\pdfoutput=1
\documentclass[12pt,a4paper]{article}
\usepackage{ifthen} % for conditional statements
\newboolean{pdflatex}
\setboolean{pdflatex}{true} % False for eps figures 

\newboolean{articletitles}
\setboolean{articletitles}{true} % False removes titles in references

\newboolean{uprightparticles}
\setboolean{uprightparticles}{false} %True for upright particle symbols

\def\paperauthors{LHCb collaboration} % Leave as is for PAPER, CONF and FIGURE
\def\paperasciititle{} % Set ASCII title here !! MAKE sure it's only ASCII characters !! 

\def\papertitle{Observation of new resonances decaying to $\jpsi\Kp$ and $\jpsi \phi$
} % Latex formatted title
\def\paperkeywords{{High Energy Physics}, {LHCb}} % Comma separated list
\def\papercopyright{\the\year\ CERN for the benefit of the LHCb collaboration} % new since 9/Apr/2018
\def\paperlicence{CC BY 4.0 licence}
\def\paperlicenceurl{https://creativecommons.org/licenses/by/4.0/}

\usepackage[top=1in, bottom=1.25in, left=1in, right=1in]{geometry}

\usepackage{microtype}
\usepackage{lineno}  % for line numbering during review
\usepackage{xspace} % To avoid problems with missing or double spaces after
\usepackage{caption} %these three command get the figure and table captions automatically small

\usepackage{graphicx}  % to include figures (can also use other packages)
\usepackage{color}
\usepackage{colortbl}
\graphicspath{{./figs/}} % Make Latex search fig subdir for figures
\usepackage{amsmath} % Adds a large collection of math symbols
\usepackage{amssymb}
\usepackage{amsfonts}
\usepackage{upgreek} % Adds in support for greek letters in roman typeset

\newcommand*\patchAmsMathEnvironmentForLineno[1]{%
\expandafter\let\csname old#1\expandafter\endcsname\csname #1\endcsname
\expandafter\let\csname oldend#1\expandafter\endcsname\csname
end#1\endcsname
 \renewenvironment{#1}%
   {\linenomath\csname old#1\endcsname}%
   {\csname oldend#1\endcsname\endlinenomath}%
}
\newcommand*\patchBothAmsMathEnvironmentsForLineno[1]{%
  \patchAmsMathEnvironmentForLineno{#1}%
  \patchAmsMathEnvironmentForLineno{#1*}%
}
\AtBeginDocument{%
\patchBothAmsMathEnvironmentsForLineno{equation}%
\patchBothAmsMathEnvironmentsForLineno{align}%
\patchBothAmsMathEnvironmentsForLineno{flalign}%
\patchBothAmsMathEnvironmentsForLineno{alignat}%
\patchBothAmsMathEnvironmentsForLineno{gather}%
\patchBothAmsMathEnvironmentsForLineno{multline}%
\patchBothAmsMathEnvironmentsForLineno{eqnarray}%
}

\usepackage{hyperxmp}

\usepackage[pdftex,
            pdfauthor={\paperauthors},
            pdftitle={\paperasciititle},
            pdfkeywords={\paperkeywords},
            pdfcopyright={Copyright (C) \papercopyright},
            pdflicenseurl={\paperlicenceurl}]{hyperref}
\usepackage[colorinlistoftodos,textsize=scriptsize]{todonotes}

\usepackage[bottom,flushmargin,hang,multiple]{footmisc}

\usepackage[all]{hypcap} % Internal hyperlinks to floats.

\usepackage{xspace} 
\usepackage{upgreek}

\def\lhcb   {\mbox{LHCb}\xspace}

\def\MagUp {\mbox{\em Mag\kern -0.05em Up}\xspace}

\ifthenelse{\boolean{uprightparticles}}%
{

 \def\Pmu         {\ensuremath{\upmu}\xspace}

 \def\Ppsi        {\ensuremath{\uppsi}\xspace}

 \def\PDelta      {\ensuremath{\Delta}\xspace}                 
 \def\PXi         {\ensuremath{\Xi}\xspace}                 
 \def\PLambda     {\ensuremath{\Lambda}\xspace}                 
 \def\PSigma      {\ensuremath{\Sigma}\xspace}                 
 \def\POmega      {\ensuremath{\Omega}\xspace}                 
 \def\PUpsilon    {\ensuremath{\Upsilon}\xspace}

 \def\PB      {\ensuremath{\mathrm{B}}\xspace}                 
                  
 \def\PD      {\ensuremath{\mathrm{D}}\xspace}

 \def\PJ      {\ensuremath{\mathrm{J}}\xspace}                 
 \def\PK      {\ensuremath{\mathrm{K}}\xspace}

 \def\Pc      {\ensuremath{\mathrm{c}}\xspace}                 
 \def\Pd      {\ensuremath{\mathrm{d}}\xspace}

 \def\Pi      {\ensuremath{\mathrm{i}}\xspace}

 \def\Ps      {\ensuremath{\mathrm{s}}\xspace}                 
                  
 \def\Pu      {\ensuremath{\mathrm{u}}\xspace}

 \def\thebaroffset{0.0em}
}
{

 \def\Pmu         {\ensuremath{\mu}\xspace}

 \def\Ppsi        {\ensuremath{\psi}\xspace}                 
                  
 \mathchardef\PDelta="7101
 \mathchardef\PXi="7104
 \mathchardef\PLambda="7103
 \mathchardef\PSigma="7106
 \mathchardef\POmega="710A
 \mathchardef\PUpsilon="7107
                  
 \def\PB      {\ensuremath{B}\xspace}                 
                  
 \def\PD      {\ensuremath{D}\xspace}

 \def\PJ      {\ensuremath{J}\xspace}                 
 \def\PK      {\ensuremath{K}\xspace}

 \def\Pc      {\ensuremath{c}\xspace}                 
 \def\Pd      {\ensuremath{d}\xspace}

 \def\Pi      {\ensuremath{i}\xspace}

 \def\Ps      {\ensuremath{s}\xspace}                 
                  
 \def\Pu      {\ensuremath{u}\xspace}

 \def\thebaroffset{0.18em}
}
\newcommand{\offsetoverline}[2][\thebaroffset]{\kern #1\overline{\kern -#1 #2}}%

\makeatletter
\ifcase \@ptsize \relax% 10pt
  \newcommand{\miniscule}{\@setfontsize\miniscule{4}{5}}% \tiny: 5/6
\or% 11pt
  \newcommand{\miniscule}{\@setfontsize\miniscule{5}{6}}% \tiny: 6/7
\or% 12pt
  \newcommand{\miniscule}{\@setfontsize\miniscule{5}{6}}% \tiny: 6/7
\fi
\makeatother

\DeclareRobustCommand{\optbar}[1]{\shortstack{{\miniscule (\rule[.5ex]{1.25em}{.18mm})}
  \\ [-.7ex] $#1$}}

\def\mup        {{\ensuremath{\Pmu^+}}\xspace}
\def\mun        {{\ensuremath{\Pmu^-}}\xspace} % muon negative (\mum is taken)

\def\uquark    {{\ensuremath{\Pu}}\xspace}

\def\dquark    {{\ensuremath{\Pd}}\xspace}

\def\squark    {{\ensuremath{\Ps}}\xspace}
\def\squarkbar {{\ensuremath{\overline \squark}}\xspace}

\def\cquark    {{\ensuremath{\Pc}}\xspace}
\def\cquarkbar {{\ensuremath{\overline \cquark}}\xspace}
\def\ccbar     {{\ensuremath{\cquark\cquarkbar}}\xspace}

\def\kaon    {{\ensuremath{\PK}}\xspace}
\def\KorKbar {\kern \thebaroffset\optbar{\kern -\thebaroffset \PK}{}\xspace}

\def\Kp      {{\ensuremath{\kaon^+}}\xspace}
\def\Km      {{\ensuremath{\kaon^-}}\xspace}

\def\D       {{\ensuremath{\PD}}\xspace}

\def\DorDbar {\kern \thebaroffset\optbar{\kern -\thebaroffset \PD}\xspace}
\def\Dz      {{\ensuremath{\D^0}}\xspace}

\def\Dp      {{\ensuremath{\D^+}}\xspace}
\def\Dm      {{\ensuremath{\D^-}}\xspace}

\def\DpDm    {\ensuremath{\Dp {\kern -0.16em \Dm}}\xspace}

\def\Dstarz  {{\ensuremath{\D^{*0}}}\xspace}

\def\Dssm    {{\ensuremath{\D^{*-}_\squark}}\xspace}

\def\B       {{\ensuremath{\PB}}\xspace}

\def\BorBbar {\kern \thebaroffset\optbar{\kern -\thebaroffset \PB}\xspace}

\def\Bd      {{\ensuremath{\B^0}}\xspace}

\def\BdorBdbar {\kern \thebaroffset\optbar{\kern -\thebaroffset \Bd}\xspace}
\def\Bu      {{\ensuremath{\B^+}}\xspace}

\def\Bp      {{\ensuremath{\Bu}}\xspace}

\def\Bs      {{\ensuremath{\B^0_\squark}}\xspace}

\def\BsorBsbar {\kern \thebaroffset\optbar{\kern -\thebaroffset \Bs}\xspace}

\def\jpsi     {{\ensuremath{{\PJ\mskip -3mu/\mskip -2mu\Ppsi}}}\xspace}

\def\Y#1S{\ensuremath{\PUpsilon{(#1S)}}\xspace}

\def\LorLbar     {\kern \thebaroffset\optbar{\kern -\thebaroffset \PLambda}\xspace}

\def\to                 {\ensuremath{\rightarrow}\xspace}

\def\AT#1     {\ensuremath{A_{\mathrm{T}}^{#1}}\xspace}           % 2

\def\C#1      {\ensuremath{\mathcal{C}_{#1}}\xspace}                       % 9
\def\Cp#1     {\ensuremath{\mathcal{C}_{#1}^{'}}\xspace}                    % 7
\def\Ceff#1   {\ensuremath{\mathcal{C}_{#1}^{\mathrm{(eff)}}}\xspace}        % 9  
\def\Cpeff#1  {\ensuremath{\mathcal{C}_{#1}^{'\mathrm{(eff)}}}\xspace}       % 7
\def\Ope#1    {\ensuremath{\mathcal{O}_{#1}}\xspace}                       % 2
\def\Opep#1   {\ensuremath{\mathcal{O}_{#1}^{'}}\xspace}                    % 7

\newcommand{\aunit}[1]{\ensuremath{\text{\,#1}}}       
\newcommand{\tev}{\aunit{Te\kern -0.1em V}\xspace}
\newcommand{\gev}{\aunit{Ge\kern -0.1em V}\xspace}
\newcommand{\mev}{\aunit{Me\kern -0.1em V}\xspace}
\newcommand{\kev}{\aunit{ke\kern -0.1em V}\xspace}
\newcommand{\ev}{\aunit{e\kern -0.1em V}\xspace}
 
\newcommand{\mevc}{\ensuremath{\aunit{Me\kern -0.1em V\!/}c}\xspace}
\newcommand{\gevc}{\ensuremath{\aunit{Ge\kern -0.1em V\!/}c}\xspace}
\newcommand{\mevcc}{\ensuremath{\aunit{Me\kern -0.1em V\!/}c^2}\xspace}
\newcommand{\gevcc}{\ensuremath{\aunit{Ge\kern -0.1em V\!/}c^2}\xspace}
\def\fb   {\ensuremath{\aunit{fb}}\xspace}
\def\invfb   {\ensuremath{\fb^{-1}}\xspace}

\newcommand{\stat}{\aunit{(stat)}\xspace}
\newcommand{\syst}{\aunit{(syst)}\xspace}

\newcommand{\chisq}{\ensuremath{\chi^2}\xspace}

\newcommand{\chisqip}{\ensuremath{\chi^2_{\text{IP}}}\xspace}

\def\gsim{{~\raise.15em\hbox{$>$}\kern-.85em
          \lower.35em\hbox{$\sim$}~}\xspace}
\def\lsim{{~\raise.15em\hbox{$<$}\kern-.85em
          \lower.35em\hbox{$\sim$}~}\xspace}

\def\sPlot{\mbox{\em sPlot}\xspace}

\def\tell1  {TELL1\xspace}
\def\ukl1   {UKL1\xspace}

\newcommand{\ie}{\mbox{\itshape i.e.}\xspace}

\def\Zcs {\ensuremath{Z_{cs}^+}\xspace}

\newcommand{\mygevc}{\ensuremath{{\mathrm{Ge\kern -0.1em V\!/}c}}\xspace}
\newcommand{\xx}{\ensuremath{\kern 0.5em }}

\newcommand{\Bpdecay}{\mbox{\ensuremath{\Bp\to\jpsi\phi\Kp}\xspace}}

\def\sPlot{\mbox{\em sPlot}\xspace}

\usepackage{multirow} % for complicated table
\usepackage{booktabs} % for complicated table
\usepackage{rotating}
\usepackage[utf8]{inputenc}

\usepackage{cite} % Allows for ranges in citations
\usepackage{mciteplus}
\usepackage{longtable} % only for template; not usually to be used in PAPERs

\newcommand{\nslj}[4]{#1{}^{#2}{\rm #3}_{#4}}

\begin{document}

%%%%%%%%%%%%%%%%%%%%%%%%%
%%%%% Title     %%%%%%%%%
%%%%%%%%%%%%%%%%%%%%%%%%%
\renewcommand{\thefootnote}{\fnsymbol{footnote}}
\setcounter{footnote}{1}

% %%%%%%% CHOOSE TITLE PAGE--------
%\onecolumn
%\input{title-LHCb-INT}
%\input{title-LHCb-ANA}
%\input{title-LHCb-CONF}
%\input{title-LHCb-FIGURE}
% ===============================================================================
% Purpose: LHCb-PAPER journal paper title page template
% Author: 
% Created on: 2010-09-25
% ===============================================================================

%%%%%%%%%%%%%%%%%%%%%%%%%
%%%%%  TITLE PAGE  %%%%%%
%%%%%%%%%%%%%%%%%%%%%%%%%
\begin{titlepage}
\pagenumbering{roman}

% Header ---------------------------------------------------
\vspace*{-1.5cm}
\centerline{\large EUROPEAN ORGANIZATION FOR NUCLEAR RESEARCH (CERN)}
\vspace*{1.5cm}
\noindent
\begin{tabular*}{\linewidth}{lc@{\extracolsep{\fill}}r@{\extracolsep{0pt}}}
\ifthenelse{\boolean{pdflatex}}% Logo format choice
{\vspace*{-1.5cm}\mbox{\!\!\!\includegraphics[width=.14\textwidth]{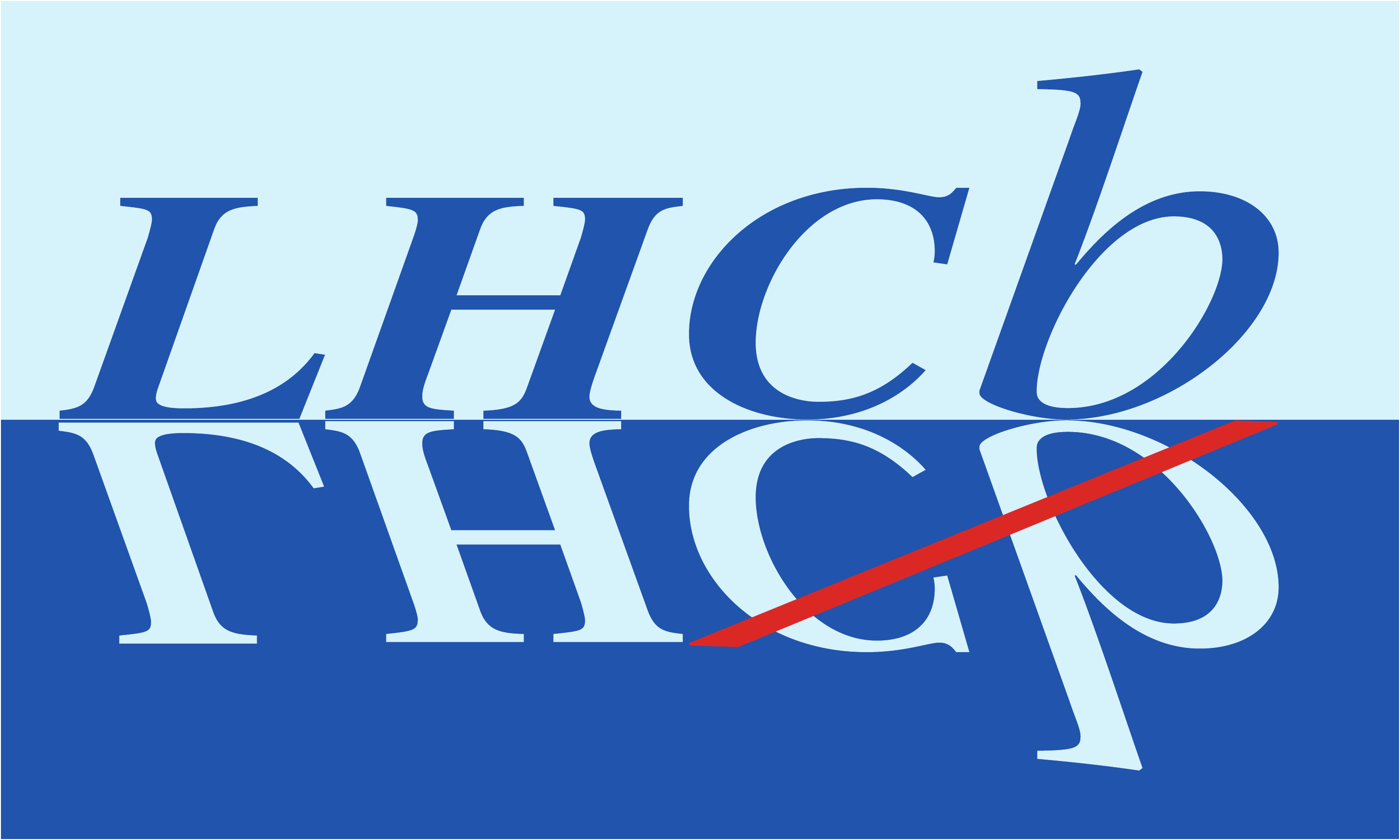}} & &}%
{\vspace*{-1.2cm}\mbox{\!\!\!\includegraphics[width=.12\textwidth]{lhcb-logo.eps}} & &}%
\\
 & & CERN-EP-2021-025 \\  % ID 
 & & LHCb-PAPER-2020-044 \\  % ID 
 & & \today \\ % Date - Can also hardwire e.g.: 23 March 2010
% not in paper \hline
\end{tabular*}

\vspace*{4.0cm}

% Title --------------------------------------------------
{\normalfont\bfseries\boldmath\huge
\begin{center}
% DO NOT EDIT HERE. Instead edit macro in main.tex to keep metadata correct
  \papertitle 
\end{center}
}

\vspace*{2.0cm}

% Authors -------------------------------------------------
\begin{center}
\paperauthors\footnote{Authors are listed at the end of this paper.}
\end{center}

\vspace{\fill}

% Abstract -----------------------------------------------
\begin{abstract}
  \noindent
  The first observation of exotic states with a new quark content $\ccbar u\squarkbar$ decaying to the $\jpsi K^+$ final state is reported with high significance from an amplitude analysis of the 
  %$\Bpdecay$ 
  $\Bp \to \jpsi \phi K^+$
  decay. The analysis is carried out using proton-proton collision data corresponding to a total integrated luminosity of 9\invfb collected by the LHCb experiment at centre-of-mass energies of $7$, $8$ and $13$\tev. The most significant state, $Z_{cs}(4000)^+$, has a mass of $4003\pm6\,^{+\,\phantom{0}4}_{-\,14}\mev$, a width of $131\pm15\pm26\mev$, and spin-parity $J^P=1^+$, where the quoted uncertainties are statistical and systematic, respectively. A new $1^+$ $X(4685)$ state decaying to the $\jpsi \phi$ final state is also observed with high significance. In addition, the four previously reported $\jpsi \phi$ states are confirmed and two more exotic states, $Z_{cs}(4220)^+$ and $X(4630)$,  are observed with significance exceeding five standard deviations.
\end{abstract}

\vspace*{2.0cm}

\begin{center}
   Published in 
   Phys. Rev. Lett. 127 (2021) 082001
%   Submitted to
%  JHEP /
%  Phys.~Rev.~D /
%  Phys.~Rev.~Lett.
%  Phys.~Lett.~B /
%  Eur.~Phys.~J.~C /
  %  Nucl.~Phys.~B /
%  Chin.~Phys.~C /
%  Nature~Physics /
%  sciPost~Physics /
%  J. Instr. /
%  Instruments 
\end{center}

\vspace{\fill}

{\footnotesize 
% Edit macro in main.tex to keep metadata correct
\centerline{\copyright~\papercopyright. \href{\paperlicenceurl}{\paperlicence}.}}
\vspace*{2mm}

\end{titlepage}

%%%%%%%%%%%%%%%%%%%%%%%%%%%%%%%%
%%%%%  EOD OF TITLE PAGE  %%%%%%
%%%%%%%%%%%%%%%%%%%%%%%%%%%%%%%%

%  empty page follows the title page ----
\newpage
\setcounter{page}{2}
\mbox{~}

%\twocolumn
% %%%%%%%%%%%%% ---------

\renewcommand{\thefootnote}{\arabic{footnote}}
\setcounter{footnote}{0}

%%%%%%%%%%%%%%%%%%%%%%%%%%%%%%%%
%%%%%  Table of Content   %%%%%%
%%%%%%%%%%%%%%%%%%%%%%%%%%%%%%%%
%%%% Uncomment if desired
%\tableofcontents
\cleardoublepage

%%%%%%%%%%%%%%%%%%%%%%%%%
%%%%% Main text %%%%%%%%%
%%%%%%%%%%%%%%%%%%%%%%%%%

\pagestyle{plain} % restore page numbers for the main text
\setcounter{page}{1}
\pagenumbering{arabic}

Charged states such as $Z_c(3900)^+$~\cite{Ablikim:2013mio,Liu:2013dau} and $Z_c(4430)^+$~\cite{Choi:2007wga, Chilikin:2013tch, LHCb-PAPER-2014-014} provide evidence for %tetraquark 
exotic states, because light quarks are required to account for the non-zero electric charge in addition to the heavy quarkonium.\footnote{Charge conjugation is implied throughout this Letter.} Previously, only the \uquark or \dquark quarks were observed to constitute the light quark content of such charged exotic states, even though the existence of a $Z_{cs}$ state as a strangeness-flavour partner of the $Z^+_c(3900)$ state had been predicted~\cite{Voloshin:2019ilw,Dias:2013qga,Chen:2013wca,Ferretti:2020ewe,Lee:2008uy}. Recently, the BESIII experiment reported a 5.3 standard deviation ($\sigma$ hereafter) observation of a threshold structure in the mass distribution of $D_s^-\Dstarz+\Dssm\Dz$ pairs produced in $e^+e^-$ annihilation as recoil against a $K^+$ meson~\cite{Ablikim:2020hsk}. 

In this Letter, the first observation of two charged $Z_{cs}^+\to\jpsi K^+$ states is reported from an updated amplitude analysis of the $\Bpdecay$ decay, as well as the observation of two more $X\to\jpsi\phi$ states. The analysis is based on the combined proton-proton ($pp$) collision data  collected using the LHCb detector in Run 1 at centre-of-mass energies $\sqrt{s}$ of 7 and 8\tev, corresponding to a total integrated luminosity of 3\invfb, and in Run 2 at $\sqrt{s}=13\tev$ corresponding to an integrated luminosity of 6\invfb.

With Run 1 data, LHCb performed the first amplitude analysis of the $\Bpdecay$ decay, 
investigating  the $\jpsi \phi$ structure~\cite{LHCb-PAPER-2016-018,LHCb-PAPER-2016-019} in addition to the kaon excitations (hereafter indicated as $K^{*+}$).
The data were described with seven $K^{*+}\to\phi K^+$ resonances,
four $X\to\jpsi\phi$ structures, and non-resonant (NR) $\phi K^+$ and $\jpsi \phi$ contributions. 
Four $X$ structures, \ie the $X(4140)$, $X(4274)$, $X(4500)$ and $X(4700)$ states were observed (the recent PDG convention labels these states as $\chi_{cJ}$ \cite{PDG2020}).
Notably, 
the $X(4140)$ width was substantially larger than previously determined~\cite{Aaltonen:2009tz,Aaltonen:2011at,Chatrchyan:2013dma}. 
Only $~3\sigma$ evidence for a $\Zcs \to\jpsi\Kp$ contribution was found~\cite{LHCb-PAPER-2016-018,LHCb-PAPER-2016-019}.

The LHCb detector is a 
single-arm forward spectrometer covering the pseudorapidity range 
$2 < \eta < 5$, 
described in detail in Refs.~\cite{LHCb-DP-2008-001,LHCb-DP-2014-002}. 
Simulation is produced with software packages
described in Refs.~\cite{Sjostrand:2007gs,LHCb-PROC-2010-056,Lange:2001uf,Allison:2006ve,*Agostinelli:2002hh}. The \mbox{$\Bp\to\jpsi(\to\mup\mun)\phi(\to \Kp\Km) \Kp$} signal candidates are first required to pass an online event selection performed by a trigger~\cite{LHCb-DP-2012-004} dedicated for selecting $\jpsi$ candidates.  
The signal decay is reconstructed by combining the $\jpsi$ candidate with three kaon candidates with a total charge of one unit. The $\phi$ candidate is selected by requiring only one of two $K^+K^-$ combinations to be consistent with the known $\phi$ mass~\cite{PDG2020} within $\pm15$\mev.\footnote{Natural units with $\hbar=c=1$ are used throughout.}

\begin{figure}[t]
\centering
\includegraphics[width=0.55\textwidth]{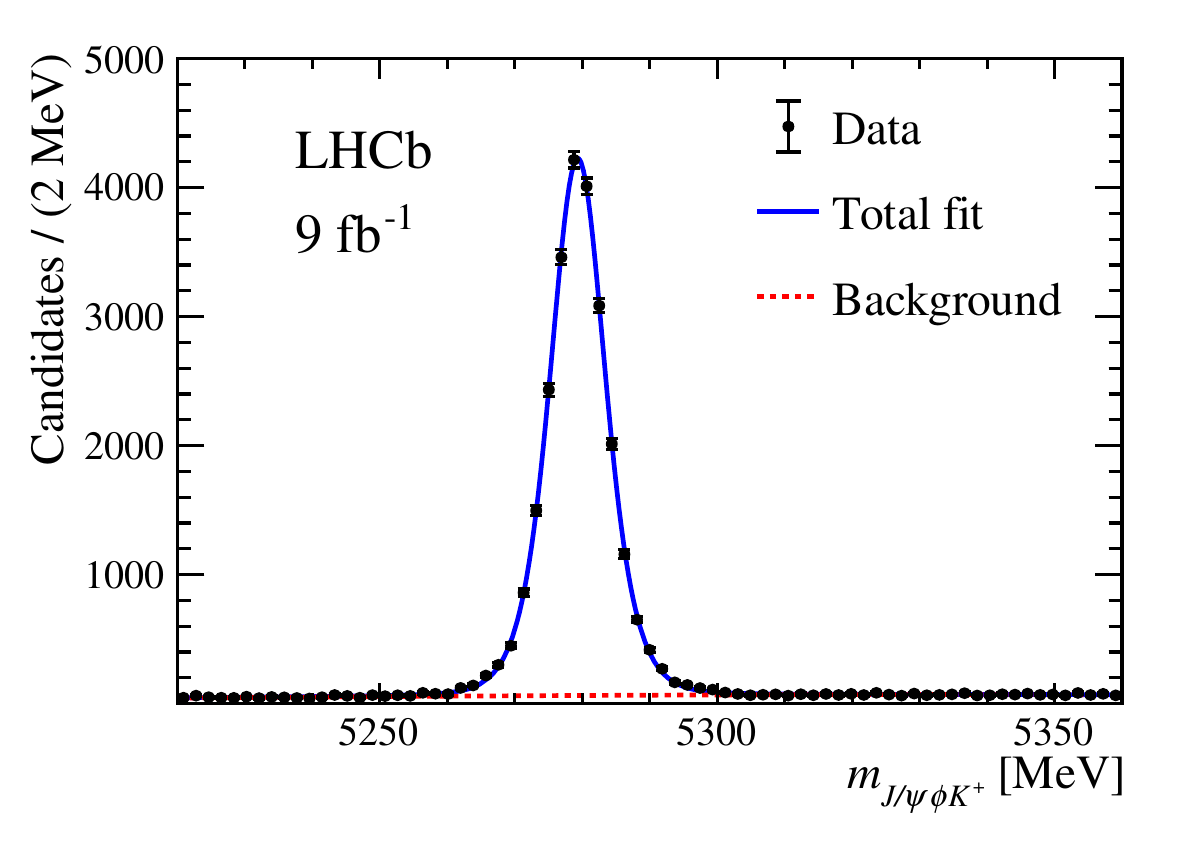}%
\caption{Invariant-mass distribution of selected \Bpdecay candidates with the fit overlaid.}
\label{massfit}
\end{figure}

The offline selection involves a loose preselection,
followed by a multivariate classifier based on a Gradient Boosted Decision Tree (BDTG)~\cite{Breiman,*Hocker:2007ht,TMVA4}. The preselection is similar to that used in Refs.~\cite{LHCb-PAPER-2016-018,LHCb-PAPER-2016-019} but the requirement on the $\chisqip$ of kaon candidates is loosened, 
where $\chisqip$ is defined as the difference in the vertex-fit \chisq of the event primary $pp$ collision vertex (PV) candidate, reconstructed with and without the particle considered. 
The BDTG response is constructed using eight variables exploring decay topology, particle momenta components transverse to the beam direction, and particle identification information (PID). 
The requirement on the BDTG response is chosen to maximise the signal significance multiplied by the purity~\cite{LHCb-PAPER-2018-045}.

The invariant-mass distribution of the $\Bp\to\jpsi\phi\Kp$ candidates is shown in Fig.~\ref{massfit}, fitted with
the signal modelled by a Hypatia function~\cite{Santos:2013gra} and the combinatorial background by a second-order polynomial function, yielding $24\,220\pm170$ signal candidates with a combinatorial-background fraction of 4.0\% within a  $\pm15$\mev signal region. The region also includes an additional $\sim2\%$ of non-$\phi$ $\Bp\to\jpsi\Kp\Km\Kp$ background candidates, which are neglected in the amplitude model but considered in the evaluation of the systematic uncertainties. The candidates in the signal region are retained for further amplitude analysis. 
Compared to the previous Run 1 analysis~\cite{LHCb-PAPER-2016-018,LHCb-PAPER-2016-019}, 
the total signal yield is $\sim6$ times larger, 
owing to a larger dataset and 
increase of 15\% in signal efficiency due to the inclusion of PID in the BDTG classifier.
The fraction of combinatorial background is almost a factor of six smaller while that of the non-$\phi$ background is unchanged. 

\begin{figure}[bt]
\centering
\includegraphics[width=0.48\textwidth]{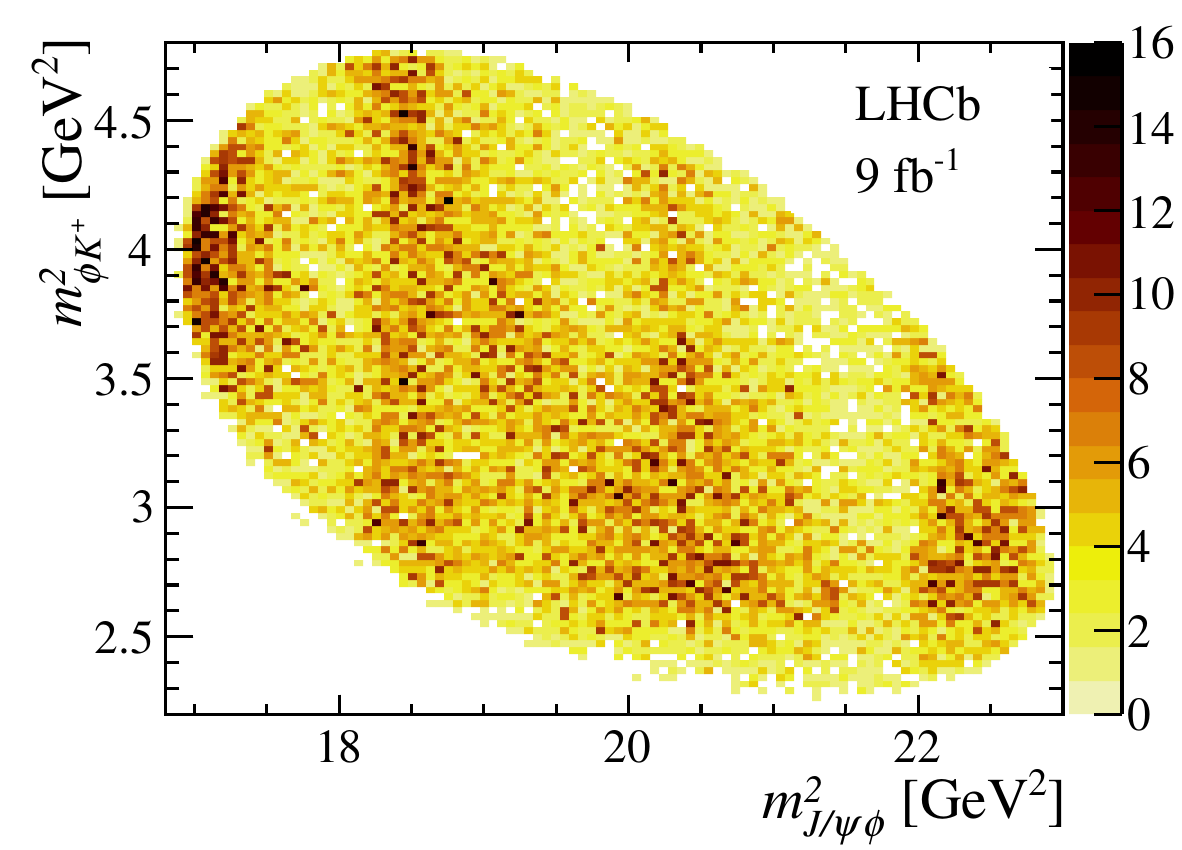}%
\includegraphics[width=0.48\textwidth]{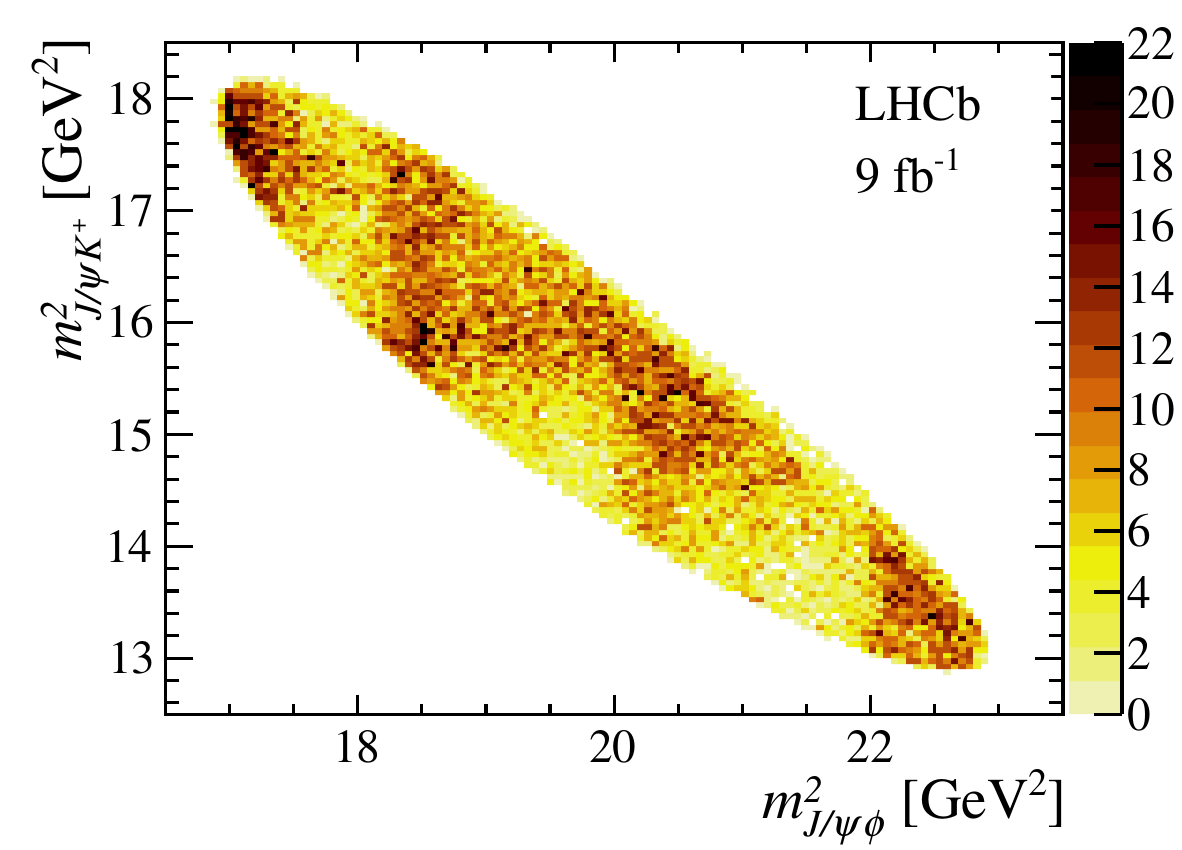}
\caption{Dalitz plots for $\Bu\to\jpsi\phi\Kp$ candidates in a region $\pm$15\mev around the $B^+$ mass peak.} 
\label{fig:dalitz}
\end{figure}

Figure~\ref{fig:dalitz} shows the Dalitz plots for $\Bp\to\jpsi\phi\Kp$ candidates in the $\Bp$ signal region.
The most apparent features are four bands in the $\jpsi\phi$ mass distribution, corresponding to the previously reported $X(4140)$, $X(4274)$, $X(4500)$ and $X(4700)$ states.
There is also a distinct band near $16\gev^2$ of the $\jpsi\Kp$  mass squared.

To investigate the resonant structures, a full amplitude fit is performed using an unbinned maximum-likelihood method.  The likelihood definition and the total probability density function (PDF), which includes a signal and a background component, are described in the previous publication~\cite{LHCb-PAPER-2016-019}. Resonance lineshapes are parametrised using the Breit-Wigner approximation. The signal $\Bp$ decay is described in the helicity formalism  by three decay chains: $K^{*+}(\to\phi K^+)\jpsi$, $X(\to\jpsi \phi) K^+$ and $Z^+_{cs}(\to \jpsi K^+)\phi$. Each chain is fully described by one mass and five angular observables. For example, the conventional $K^{*+}$ chain has the following six observables $\Phi\equiv(m_{\phi K}, \theta_{K^*}, \theta_{\jpsi}, \theta_{\phi}, \Delta\varphi_{K^*, \jpsi}, \Delta\varphi_{K^*,\phi})$,
where $\theta$ denotes the helicity angles, and $\Delta\varphi$ the angles between two decay planes. Due to the non-scalar final-state particles ($\mu^+$ and $\mu^-$), an azimuthal angle $\alpha_\mu^i$ is required to align the helicity frames of $\mu^+$ and $\mu^-$ between the chain $i$ and the reference $K^{*+}$ chain\cite{Mizuk:2008me, Chilikin:2013tch, LHCb-PAPER-2014-014}.

\begin{figure}[bt]
\centering
\includegraphics[width=1.0\textwidth]{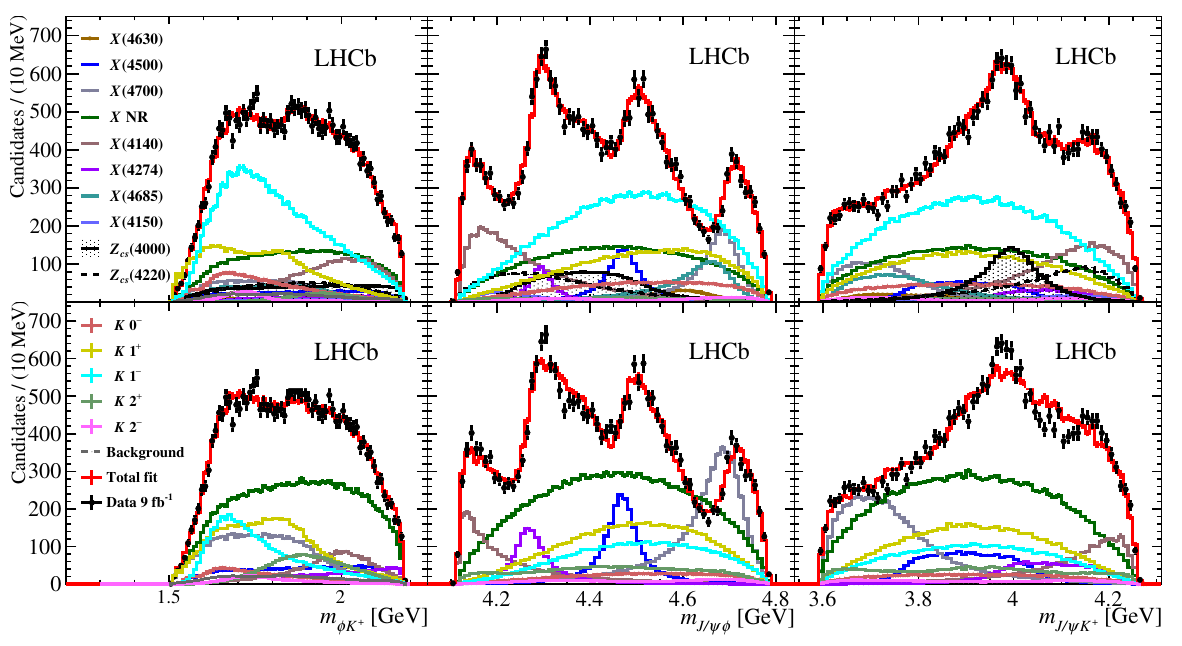}
\caption{Distributions of $\phi\Kp$ (left), $\jpsi\phi$ (middle) and  $\jpsi\Kp$ (right) invariant masses for the $\Bp\to\jpsi\phi\Kp$ candidates (black data points) compared with the fit results (red solid lines) of the default model (top row) and the Run 1 model (bottom row).
} 
\label{fig:default_model}
\end{figure}

The model used in the previous study (Run 1 model) is first tested. Due to the increased sample size, the model requires improvements (see Fig.~\ref{fig:default_model} bottom row). 
Additional $K^{*+}$, $X$ and possible $\Zcs$ states 
are added until no further state with a significance larger than $5\sigma$ improves the overall fit. 
In total, nine $K^{*+}$, seven $X$, two $\Zcs$, and one $\jpsi\phi$ NR components are taken as the default model, as listed in Table~\ref{tab:fit}. 
The nine $K^{*+}$ states are all those with spin-parity $J\le 2$ and mass below 2\gev, 
which are predicted by the relativistic potential model~\cite{Godfrey:1985xj}, 
and kinematically allowed, including three resonances with poles just below the $\phi K^+$ mass threshold. 
All components previously used in the Run 1 model are included, but the $J^P=1^+$ NR $\phi K^+$, and the broad $0^-$ state, are replaced by the upper tails of $K_1(1400)$ and $K(1460)$ resonances, respectively. The newly added components are: the upper tail of $1^-$ $K^*(1410)$ resonance, $2^-$ $X(4150)$, $1^+$ $X(4685)$, $1^-$ $X(4630)$, $1^+$ $Z_{cs}(4000)^+$ and $Z_{cs}(4220)^+$ states.

\begin{figure}[!tbp]
\centering

\includegraphics[width=0.8\textwidth]{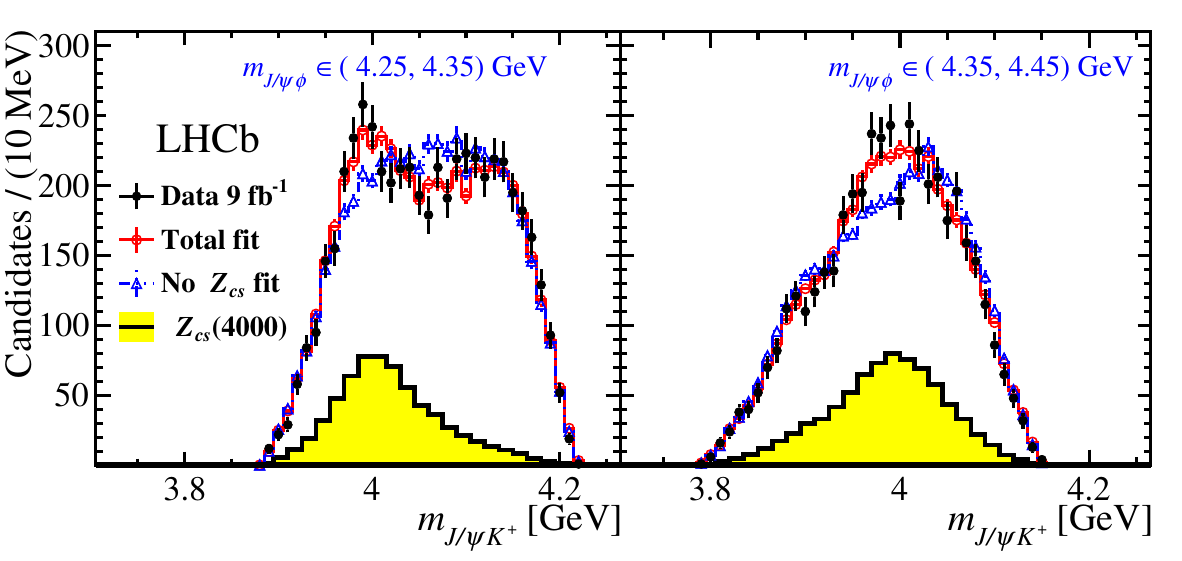}%

\caption{Projections of the fits with the default model, performed in the full phase-space, onto $m_{\jpsi\Kp}$ in two slices of $m_{\jpsi\phi}$ with and without the $1^+$ $\Zcs$ states. The narrow $\Zcs$ state at 4 \gev is evident.}
\label{fig:fitmjpsik}
\end{figure}

\begin{table}[hbtp]
\begin{center}
\caption{Fit results from the default amplitude model. The significances are evaluated accounting for total (statistical) uncertainties. The listed masses and widths without uncertainties are taken from PDG~\cite{PDG2020} and are fixed in the fit. The listed world averages of the two $K_2$ and $K^*(1680)$ resonances do not contain the contributions from the previous LHCb Run 1 results.}\label{tab:fit}
\vskip -0.2cm
\def\arraystretch{1.2}
\begin{tabular}{c cccr@{\:$\pm$\:}lr@{\:$\pm$\:}lr@{\:$\pm$\:}l}
   $J^{P}$ & \multicolumn{2}{c}{Contribution} &Significance\,[$\times \sigma$]&            \multicolumn{2}{c}{$M_0$\,[MeV]}   & \multicolumn{2}{c}{$\Gamma_0$\,[MeV]}      & \multicolumn{2}{c}{FF\,[\%]}          \\
\hline
\multirow{3}{*}{$1^+$} &
$\nslj{2}{1}{P}{1}$   &  $K(1^+)$            & 4.5 (4.5)  &  $1861$&$10\,_{-\,46}^{+\,16} $  & $149$&$41\,_{-\,\phantom{0}23}^{+\,231}$ & \multicolumn{2}{c}{}\\
   &
$\nslj{2}{3}{P}{1}$   &  $K^{\prime}$($1^+$) & 4.5 (4.5)  &  $1911$&$37\,_{-\,\phantom{0}48}^{+\,124} $  &  $276$&$50\,_{-\,159}^{+\,319}$   &     \multicolumn{2}{c}{}  \\
   &
$\nslj{1}{3}{P}{1}$   &  $K_1(1400)$         & 9.2 (11)   & \multicolumn{2}{c}{1403} &            \multicolumn{2}{c}{174} &  $15$&$3\,_{-\,11}^{+\,\phantom{0}3}$    \\
\hline

\multirow{2}{*}{$2^-$} &
$\nslj{1}{1}{D}{2}$   &  $K_2 (1770)$   & 7.9 (8.0)   & \multicolumn{2}{c}{1773}   & \multicolumn{2}{c}{186}   & \multicolumn{2}{c}{}    \\
& $\nslj{1}{3}{D}{2}$   &  $K_2(1820)$    & 5.8 (5.8)   & \multicolumn{2}{c}{1816}   & \multicolumn{2}{c}{276}   &  \multicolumn{2}{c}{}   \\       

\hline
\multirow{2}{*}{$1^-$} &
$\nslj{1}{3}{D}{1}$   &  $K^*(1680)$    & 4.7 (13)    & \multicolumn{2}{c}{1717}  & \multicolumn{2}{c}{322}   &  $14$&$2\,_{-\,\phantom{0}8}^{+\,35}$     \\  
   &  $\nslj{2}{3}{S}{1}$   &  $K^*(1410)$    & 7.7 (15)    & \multicolumn{2}{c}{1414}  & \multicolumn{2}{c}{232}   &  $38$&$5\,_{-\,17}^{+\,11}$   \\
\hline

\multirow{1}{*}{$2^-$} &
$\nslj{2}{3}{P}{2}$   &  $K^*_2(1980)$  & 1.6 (7.4)   &  $1988$&$22\,_{-\,\phantom{0}31}^{+\,194} $   &   $318$&$82\,_{-\,101}^{+481}$   & $2.3$&$0.5\pm0.7$        \\  
\hline 
\multirow{1}{*}{$0^-$} &
$\nslj{2}{1}{S}{0}$   &  $K(1460)$      & 12 (13)     &  \multicolumn{2}{c}{1483}  &     \multicolumn{2}{c}{336} & $10.2$&$1.2\,_{-\,3.8}^{+\,1.0}$    \\  

\hline%\hline
\multirow{1}{*}{$2^-$} & 
   &$X(4150)$            & 4.8 (8.7) & $4146$&$18\pm33$ & $135$&$28\,_{-\,30}^{+\,59}$   & $2.0$&$0.5\,_{-\,1.0}^{+\,0.8}$ \\
\hline
\multirow{1}{*}{$1^-$} &
   &$X(4630)$            & 5.5 (5.7) & $4626$&$ 16\,_{-\,110}^{+\,\phantom{0}18}$ & $174$&$27\,_{-\,\phantom{0}73}^{+\,134}$  & $2.6$&$0.5\,_{-\,1.5}^{+\,2.9}$ \\
\hline

 \multirow{3}{*}{$0^+$} &
 &$X(4500)$           & 20 (20) & $4474$&$3 \pm3 $ & $77$&$6\,_{-\,\phantom{0}8}^{+\,10}$ &  $5.6$&$0.7\,_{-\,0.6}^{+\,2.4}$ \\
 &  & $X(4700)$           & 17 (18) & $4694$&$ 4 \,_{-\,\phantom{0}3}^{+\,16}$ & $87$&$8\,_{-\,\phantom{0}6}^{+\,16}$ &  $8.9$&$1.2\,_{-\,1.4}^{+\,4.9}$ \\
 &  & ${\rm NR}_{J/\psi \phi}$  & 4.8 (5.7) & \multicolumn{2}{c}{}& \multicolumn{2}{c}{}&$28$&$8\,_{-\,11}^{+\,19}$ \\
\hline
 \multirow{3}{*}{$1^+$} &
 &$X(4140)$           & 13 (16) & $4118$&$ 11 \,_{-\,36}^{+\,19}$ & $162$&$21\,_{-\,49}^{+\,24}$ & $17$&$3\,_{-\,6}^{+\,19}$ \\
   & &$X(4274)$           & 18 (18) & $4294$&$4 \,_{-\,6}^{+\,3}$  & $53$&$5\pm5$   & $2.8$&$0.5\,_{-\,0.4}^{+\,0.8}$ \\
   & &$X(4685)$           & 15 (15) & $4684$&$ 7\,_{-\,16}^{+\,13}$  & $126$&$15 \,_{-\,41}^{+\,37}$ & $7.2$&$1.0 \,_{-\,2.0}^{+\,4.0}$ \\
\hline%\hline
  \multirow{2}{*}{$1^+$} & 
  &$Z_{cs}(4000)$          & 15 (16) & $4003$&$ 6\,_{-\,14}^{+\,\phantom{0}4}$  & $131$&$15\pm26$  & $9.4$&$2.1\pm3.4$ \\
   &  &$Z_{cs}(4220)$          & 5.9 (8.4) & $4216$&$24\,_{-\,30}^{+\,43}$ & $233$&$52\,_{-\,73}^{+\,97}$ & $10$&$4\,_{-\,\phantom{0}7}^{+\,10}$ \\
\end{tabular}
\end{center}
\end{table}

Figure~\ref{fig:default_model} shows the invariant mass distributions for all pairs of final state particles of the $\Bp\to\jpsi\phi\Kp$ decay with fit projections from the amplitude analysis overlaid,
for both the default model and the Run 1 model.
The fit results are summarised in Table~\ref{tab:fit},
including mass, width, fit fraction (FF), and significance of each component.
The masses and widths of the four $X$ states studied using the \lhcb Run 1 sample only are consistent with the previous measurements~\cite{LHCb-PAPER-2016-018,LHCb-PAPER-2016-019}.
The significance of each component is evaluated by assuming that the change of twice the log-likelihood between the default fit and the fit without this component
follows a $\chi^2$ distribution. The corresponding  number of degrees of freedom is equal to the reduction in the number of free parameters multiplied by a factor of two (one) when the mass and width of the component are floated (fixed) in the fit, which accounts for the look-elsewhere effect~\cite{LEE,LHCb-PAPER-2016-019}, as validated by  pseudoexperiments. 
Figure~\ref{fig:fitmjpsik} shows the $m_{\jpsi\Kp}$ distributions in two slices of $m_{\jpsi\phi}$,
which demonstrate the need for the narrower $Z_{cs}(4000)^+$ state.  Including the $1^+$ $\Zcs$ states improves the $\chi^2/{\rm nbin}$ from 84/35 to 43/35 (left slice), and from 79/37 to 32/37 (right slice), where nbin is the number of non-zero bins.

The spin and parity of each exotic state is probed 
by testing alternative $J^P$ hypotheses and comparing the fit likelihood values \cite{LHCb-PAPER-2016-019}.
The $J^P$ assignments for the previously reported four $X$ states are confirmed with high significance. 
A $1^+$ assignment is favoured for the new $X(4685)$ state  with also high significance, 
but the quantum numbers of the $X(4150)$ and $X(4630)$ are less well determined.
The best hypothesis for the $X(4630)$ state is $1^-$ over $2^-$ at a $3\sigma$ level. The other hypotheses are ruled out by more than $5\sigma$. The fit prefers $2^-$ for the $X(4150)$ state by more than $4\sigma$. 
The narrower $Z_{cs}(4000)^+$ state is determined to be $1^+$ with high significance. 
The broader $Z_{cs}(4220)^+$ state could be $1^+$ or $1^-$, with a $2\sigma$ difference in favour of the first hypothesis. Other spin-parity assignments are ruled out at $4.9\sigma$ level.

Systematic uncertainties are estimated for the masses,
widths, 
and fit fractions of all states. %of the exotic states,
To probe the effects from the neglected $B^+\to\jpsi K^+K^-K^+$ non-$\phi$ contributions,
the $\phi$ mass window is changed from $\pm15\mev$ to $\pm7\mev$, and alternatively this background is subtracted  using the \sPlot technique\cite{Pivk:2004ty}. 
The Blatt-Weisskopf barrier~\cite{LHCb-PAPER-2016-019} hadron size is varied between 1.5 and 4.5\,GeV$^{-1}$.
The default NR $0^+$ $\jpsi\phi$ representation is changed from a constant to a linear polynomial.
Additional $1^+$ or $2^+$ NR $\jpsi\phi$ contributions are also included. 
The smallest allowed orbital angular momentum in the resonance function is varied.
For the $X(4140)$, which peaks near the $\jpsi\phi$ threshold,  
the  Flatt\'e model\cite{Flatte:1976xu} is used instead of 
the Breit-Wigner amplitude.
A simplified one-channel K-matrix model~\cite{PDG2020} is used to describe various $K^*$ resonances instead of the sum of Breit-Wigner amplitudes. 
Two-channel K-matrix models have also been tried for the $\nslj{2}{1}{P}{1}$ and $\nslj{2}{3}{P}{1}$ $K^*$ states with the coupled-channel thresholds opening up near 1.75\gev, with an insignificant improvement to the description of the $m_{\phi K}$ distribution.
To cover the full range of $K^{*+}$ resonances predicted in the allowed $\phi K^+$ mass range,  
an extended model is tested by adding five more $K^{*+}$ resonances with mass above 2\gev \cite{Godfrey:1985xj}.
The presence of an extra $X$ state contribution, with $J$ from 0 to 2, to the extended model is also checked.
The difference between the results obtained from assigning $1^+$ or $1^-$ hypotheses to the $Z_{cs}(4220)^+$ is taken as a systematic uncertainty.
The mass-dependent width in the denominator of the Breit-Wigner function for the $K^{*+}$ resonances is calculated with  the lightest  allowed channel ($\pi K$ for natural spin-parity resonances and $\omega K$ for others) instead of $\phi K$. 

The maximum deviation among the modelling uncertainties discussed above is summed in quadrature with the additional sources, %listed below
including the uncertainties due to the fixed masses and widths of the known $K^{*+}$ resonances,
mismodelling of $\chi^2_{\rm IP}$ of the $\Bp$ candidate,
background PDF model shape and fractions,
and the finite size of the simulation samples.
For the $Z_{cs}(4000)^+$ state, the largest systematic contribution is due to the $J^P$ hypotheses of the $Z_{cs}(4220)^+$ state. The summary of fit results, including the systematic uncertainties, is listed in Table~\ref{tab:fit}. The smallest significance found when varying each of sources is taken as the significance accounting for systematic uncertainty.

\begin{figure}[!bt]
\centering
\includegraphics[width=0.4\textwidth]{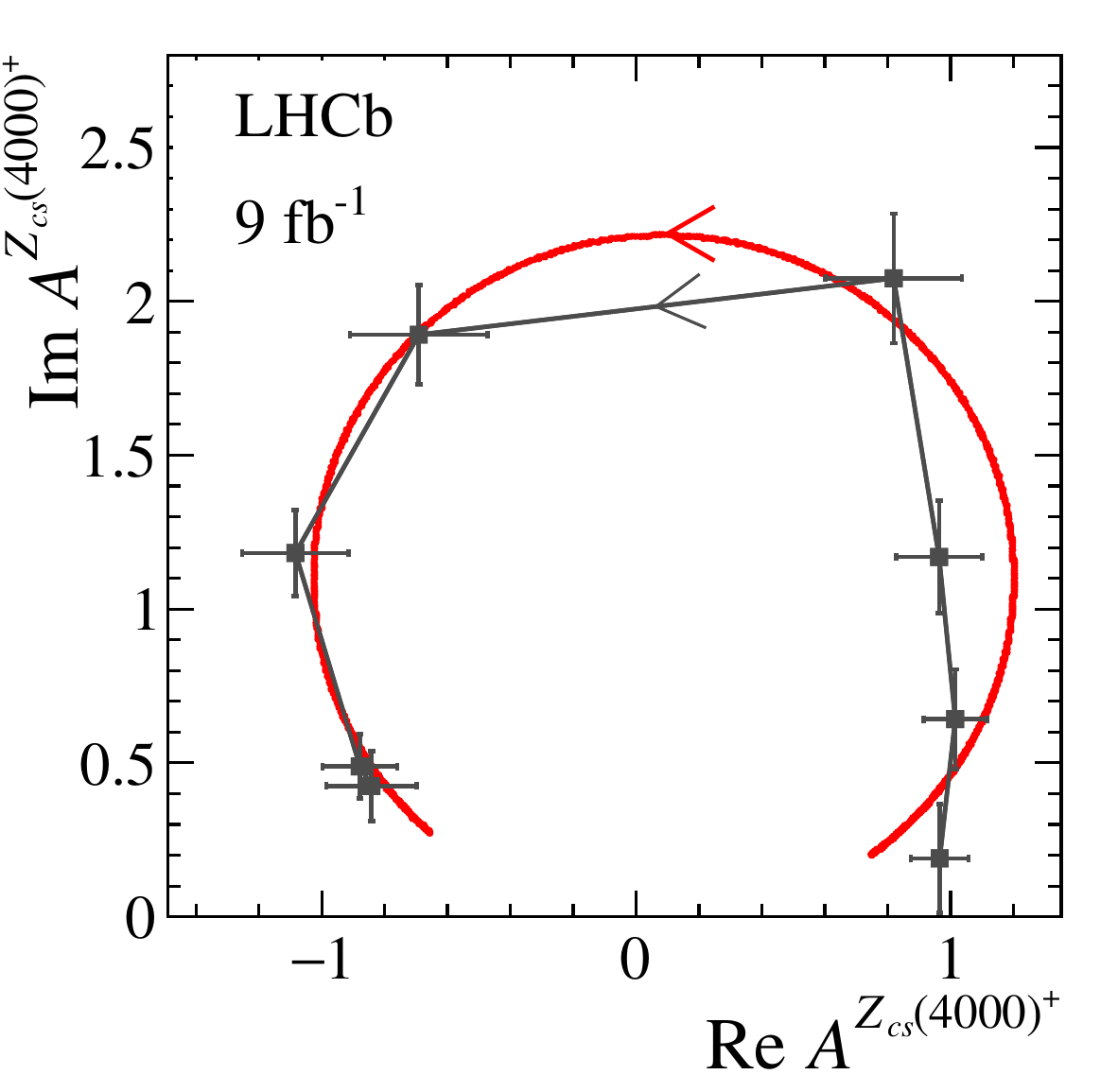}
\caption{Fitted values of the $Z_{cs}(4000)^+$ amplitude in eight $m_{\jpsi\Kp}$ intervals,
shown on an Argand diagram (black points).
The red curve represents the expected Breit-Wigner behaviour between $-1.4\Gamma_0$ to $1.4\Gamma_0$ around the $Z_{cs}(4000)^+$ mass.
} 
\label{fig:Argand}
\end{figure}

Further evidence for the resonant character of $Z_{cs}(4000)^+$ is observed in Fig.~\ref{fig:Argand}, showing the evolution of the complex amplitude on the Argand diagram, obtained with the same
method as previously reported for the  $Z_c(4430)^-$ state~\cite{LHCb-PAPER-2014-014}.
The magnitude and phase have approximately circular evolution with $m_{\jpsi\Kp}$ in the counter-clockwise direction,
as expected for a resonance.

The BESIII experiment reported observation of a $Z_{cs}(3985)^-$ resonance. Its mass $3982.5\,_{-\,2.6}^{+\,1.8}\stat\pm2.1\syst$\mev is consistent with the $1^+$ $Z_{cs}(4000)^+$ state observed in this analysis, but with significantly narrower width $12.8\,_{-\,4.4}^{+\,5.3}\stat\pm3.0\syst$\mev. When fixing the mass and width of this state to the nominal BESIII result in the amplitude fit to our data, twice the log-likelihood is worse by 160 units. The narrower width is also not supported by an alternative Flatt\'e model with parameters obtained from our data. Therefore,
there is no evidence that the $Z_{cs}(4000)^+$ state observed here is the same as the $Z_{cs}(3985)^-$ state observed by BESIII.

In conclusion, an improved full amplitude analysis of the $\Bpdecay$ decay is performed using 6 times larger signal yield than previously analyzed \cite{LHCb-PAPER-2016-018}.
A relatively narrow $Z_{cs}(4000)^+$ state decaying to $\jpsi\Kp$ with mass $4003\pm6\stat\,_{-\,14}^{+\,\phantom{0}4}\syst\mev$ and width \mbox{$131\pm15\stat\pm26\syst\mev$} is observed with large significance. 
Its spin-parity is determined to be $1^+$ also with high significance. 
A quasi-model-independent representation of the $Z_{cs}(4000)^+$ contribution in the fit shows a phase change in the amplitude consistent with that of a resonance.
A broader $1^+$ or $1^-$ $Z_{cs}(4220)^+$ state is also required at $5.9\sigma$. 
This is the first observation of states with hidden charm and strangeness that decay to the $\jpsi \Kp$ final state. 
The four $X$ states decaying to $\jpsi\phi$ observed in the Run 1 analysis \cite{LHCb-PAPER-2016-018} are confirmed with higher significance, together with their quantum number assignments. 
An additional $1^{+}$ $X(4685)$ state is observed with relatively narrow width (about 125\mev) 
with high significance. 
A new $X(4630)$ state is observed with a $5.5\sigma$ significance, with preferred $1^{-}$ over $2^-$ spin-parity assignment at $3\sigma$ level, and the other $J^{P}$ hypotheses rejected at $5\sigma$. This constitutes the first observation of exotic states with a new quark content  $\ccbar u\squarkbar$ decaying to the $\jpsi K^+$ final state.

\section*{Acknowledgements}
\noindent We express our gratitude to our colleagues in the CERN
accelerator departments for the excellent performance of the LHC. We
thank the technical and administrative staff at the LHCb
institutes.
We acknowledge support from CERN and from the national agencies:
CAPES, CNPq, FAPERJ and FINEP (Brazil); 
MOST and NSFC (China); 
CNRS/IN2P3 (France); 
BMBF, DFG and MPG (Germany); 
INFN (Italy); 
NWO (Netherlands); 
MNiSW and NCN (Poland); 
MEN/IFA (Romania); 
MSHE (Russia); 
MICINN (Spain); 
SNSF and SER (Switzerland); 
NASU (Ukraine); 
STFC (United Kingdom); 
DOE NP and NSF (USA).
We acknowledge the computing resources that are provided by CERN, IN2P3
(France), KIT and DESY (Germany), INFN (Italy), SURF (Netherlands),
PIC (Spain), GridPP (United Kingdom), RRCKI and Yandex
LLC (Russia), CSCS (Switzerland), IFIN-HH (Romania), CBPF (Brazil),
PL-GRID (Poland) and NERSC (USA).
We are indebted to the communities behind the multiple open-source
software packages on which we depend.
Individual groups or members have received support from
ARC and ARDC (Australia);
AvH Foundation (Germany);
EPLANET, Marie Sk\l{}odowska-Curie Actions and ERC (European Union);
A*MIDEX, ANR, Labex P2IO and OCEVU, and R\'{e}gion Auvergne-Rh\^{o}ne-Alpes (France);
Key Research Program of Frontier Sciences of CAS, CAS PIFI, CAS CCEPP, 
Fundamental Research Funds for the Central Universities, 
and Sci. \& Tech. Program of Guangzhou (China);
%Thousand Talents Program, and Sci. \& Tech. Program of Guangzhou (China);
RFBR, RSF and Yandex LLC (Russia);
GVA, XuntaGal and GENCAT (Spain);
the Leverhulme Trust, the Royal Society
 and UKRI (United Kingdom).

\addcontentsline{toc}{section}{References}
%\setboolean{inbibliography}{true}
\bibliographystyle{LHCb}
\bibliography{main,standard,LHCb-PAPER,LHCb-CONF,LHCb-DP,LHCb-TDR,Pentaquark,MyBib}

%\clearpage
%\section*{PRL justification}
%Exotic states, with quark configurations other than three quarks or quark-antiquark,
%can reveal new or hidden aspects of the dynamics of the strong interactions.  
%An improved full amplitude analysis of $B^+ \to J/\psi 
%\phi K^+$ decays is performed using 6 times larger signal yield than previously analyzed.
%We report the first observation of two exotic states with a new quark content $c\bar{c} u \bar{s}$ decaying to the $J/\psi K^+$ final state. Two new $X \to J/\psi \phi$ states are also observed with high significance. Four $X$ states previously observed by LHCb are also confirmed.

\newpage
% LHCb collaboration author list
% Data extracted on February 25th, 2021 at 6:02pm for paper reference LHCb-PAPER-2020-044
\centerline
{\large\bf LHCb collaboration}
\begin
{flushleft}
\small
R.~Aaij$^{32}$,
C.~Abell{\'a}n~Beteta$^{50}$,
T.~Ackernley$^{60}$,
B.~Adeva$^{46}$,
M.~Adinolfi$^{54}$,
H.~Afsharnia$^{9}$,
C.A.~Aidala$^{85}$,
S.~Aiola$^{25}$,
Z.~Ajaltouni$^{9}$,
S.~Akar$^{65}$
% ...

{\footnotesize \it

% ...
$^{9}$Universit{\'e} Clermont Auvergne, CNRS/IN2P3, LPC, Clermont-Ferrand, France\\
% ...
$^{25}$INFN Sezione di Milano, Milano, Italy\\
% ...
$^{32}$Nikhef National Institute for Subatomic Physics, Amsterdam, Netherlands\\
% ...
$^{46}$Instituto Galego de F{\'\i}sica de Altas Enerx{\'\i}as (IGFAE), Universidade de Santiago de Compostela, Santiago de Compostela, Spain\\
% ...
$^{50}$Physik-Institut, Universit{\"a}t Z{\"u}rich, Z{\"u}rich, Switzerland\\
% ...
$^{54}$H.H. Wills Physics Laboratory, University of Bristol, Bristol, United Kingdom\\
% ...
$^{60}$Oliver Lodge Laboratory, University of Liverpool, Liverpool, United Kingdom\\
% ...
$^{65}$University of Cincinnati, Cincinnati, OH, United States\\
% ...
$^{68}$Syracuse University, Syracuse, NY, United States\\
% ...
$^{85}$University of Michigan, Ann Arbor, United States, associated to $^{68}$\\
\bigskip
% ...
\medskip
}
\end{flushleft}

\end{document}